\documentclass[12pt,twoside]{article}
\usepackage[mathscr]{eucal}
\usepackage{amsmath,amsfonts,amssymb,amsthm,mathabx,empheq}
\usepackage{times}
\usepackage{pdfsync}
\usepackage{cite}
\usepackage{url}
\usepackage{hyperref}
\usepackage{tensor}
\usepackage{color}
\usepackage{multicol}
\usepackage{bbold}

\advance\textheight\topskip
\allowdisplaybreaks[1]

\newcommand{\p}{\partial}

\newcommand{\be}{\begin{equation}}
\newcommand{\ee}{\end{equation}}
\newcommand{\bea}{\begin{eqnarray}}
\newcommand{\eea}{\end{eqnarray}}

\newcommand{\nn}{\nonumber}

\allowdisplaybreaks[1]

\DeclareFontFamily{OT1}{rsfs}{} \DeclareFontShape{OT1}{rsfs}{m}{n}{
<-7> rsfs5 <7-10> rsfs7 <10-> rsfs10}{}
\DeclareMathAlphabet{\mycal}{OT1}{rsfs}{m}{n}

\hypersetup{
colorlinks=true,
linktoc=page,    
linkcolor=blue,
citecolor=blue,
urlcolor=blue,
colorlinks=true,
}

\begin{document}
\title{Notes on no black hole theorem in three dimensional gravity}

\author{Pujian Mao and Weicheng Zhao}

\date{}

\def\mytitle{Notes on no black hole theorem in three dimensional gravity}

\addtolength{\headsep}{4pt}

\begin{centering}

  \vspace{1cm}

  \textbf{\large{\mytitle}}

  \vspace{1.5cm}

  {\large Pujian Mao and Weicheng Zhao }

\vspace{0.5cm}

\begin{minipage}{.9\textwidth}\small \it  \begin{center}
     Center for Joint Quantum Studies and Department of Physics,\\
     School of Science, Tianjin University, 135 Yaguan Road, Tianjin 300350, China
 \end{center}
\end{minipage}

\vspace{0.3cm}

\end{centering}

\vspace{1cm}

\begin{center}
\begin{minipage}{.9\textwidth}
\textsc{Abstract}: In this paper, we revisit the no black hole theorem in three dimensional (3D) gravity in the Newman-Penrose formalism in a generalized Newman-Unti gauge. After adapting the well established 4D NP formalism and its gauge and coordinates system to 3D gravity, we show that the no black hole theorem is manifest in the NP equations. We further study in detail the horizon properties of the 3D charged rotating solutions and confirm that a black hole solution requires a negative cosmological constant.

\end{minipage}
\end{center}

\thispagestyle{empty}

\section{Introduction}

Einstein gravity in three dimensions has no propagating degree of freedom. Hence, it is very surprising that three dimensional Einstein gravity with negative cosmological constant admits black hole solutions \cite{Banados:1992wn}, which has received
significant attentions from the holographic perspective, see, e.g., \cite{Brown:1986nw,Strominger:1997eq,Aharony:1999ti}. It is shown that the 3D Ba\~{n}ados-Teitelboim-Zanelli (BTZ) black hole is obtained by the identification of points of anti-de Sitter space \cite{Banados:1992gq}. Then an intensely curious point is that if a similar construction can arise black hole solution for the case of a positive cosmological constant or the vanishing cosmological constant case. This question was perfectly answered by Ida with a no black hole theorem in 3D \cite{Ida:2000jh}. The proof in \cite{Ida:2000jh} can be summarized as follows. Assuming that there is a black hole in the sense of the existence of an apparent horizon in the spacetime. By definition, the expansion of a null congruence of future directed ingoing null vector orthogonal to the horizon is always positive in $(+,-,-)$ signature. While the expansion for the outgoing null vector changes sign at the horizon and should be negative outside the horizon. However, the author of \cite{Ida:2000jh} found a particular direction to deform the apparent horizon. The expansion of the outgoing null vector is positive at the deformed horizon in case that the stress tensor satisfies the dominant energy condition. That means there is a trapped region outside the apparent horizon which is in contradiction with the assumption of the existence of an apparent horizon.

In this work, we revisit the 3D no black hole theorem in the context of the Newman-Penrose (NP) formalism \cite{Newman:1961qr}. The adapted 3D NP formalism consists of two null basis vectors and one spacelike basis vector. The advantage of the NP formalism is that one can simply choose the two null basis vectors as the outgoing and ingoing null direction, which are orthogonal to the horizon, to investigate the horizon property. We construct the coordinates system in the well established Newman-Unti (NU) gauge \cite{Newman:1962cia} with slight generalization. We find that the 3D no black hole theorem is simply encoded in the near horizon solution space. There is no need to introduce the particular direction to show that it is still a trapped region outside the horizon. As a concrete example, we investigate in details of the horizon property of the 3D charged rotating solution \cite{Clement:1993kc,Martinez:1999qi}. Our results show that a black hole requires negative cosmological constant. For the positive and vanishing cosmological constant cases, there is a cosmological horizon.

This paper is organized as follows. In section \ref{2}, we first adapt the NP formalism to 3D and introduce the generalized NU gauge for the 3D NP system. We show that the 3D no black hole theorem is simply encoded in the NP equations in a transparent manner. Parallelly, we review the derivation of \cite{Ida:2000jh} with adaptions to our conventions. Then, we present the full near horizon solution space of 3D pure Einstein gravity. The solution space is in a closed form. In section \ref{3}, we transform the 3D charged rotating solution in the NU gauge. Section \ref{4} presents the detailed analysis of the horizon property of the 3D charged rotating solution in the cases of a negative, vanishing, and positive cosmological constant. In section \ref{5}, we comment on the flat limit of the 3D charged rotating solution and show that the spinning parameter of the solution with vanishing cosmological constant is simply introduced by a pure coordinates transformation. We conclude in the last section.


\section{NP formalism and no black hole theorem in 3D}
\label{2}

\subsection{3D NP formalism}

We follow the construction in \cite{Barnich:2016lyg} to adapt 3D theory in an NP-like formalism. Note that we will take the standard NP signature $(+,-,-)$ in this work. The tetrad system consists of two null vectors $n$ and $n'$ and a spacelike vector $m$. The basis vectors have the orthogonality conditions
\be
n\cdot m=n'\cdot m=0
\ee
and are normalized as
\be
n\cdot n'=1,\;\;\;\;m\cdot m=-1.
\ee
The spacetime metric is obtained from
\be\label{metric}
g_{\mu\nu}=n_\mu n'_\nu + n'_\mu n_\nu - m_\mu m_\nu .
\ee
The directional operators are defined as
\be
D=n^\mu \nabla_\mu, \qquad D'=n'^\mu\nabla_\mu, \qquad \delta=m^\mu\nabla_\mu.
\ee
Below, we list all the quantities and relations for the 3D NP system.

\noindent{\bf Spin coefficients}

\noindent
Definition:
\begin{align}
& \kappa=\Gamma_{311}=m\cdot D n,\qquad \epsilon=\Gamma_{211}=n'\cdot D n,\qquad \tau'=\Gamma_{321}=m\cdot D n', \nn \\
& \tau=\Gamma_{312}=m\cdot D' n,\qquad \epsilon'=\Gamma_{122}=n\cdot D' n',\qquad \kappa'=\Gamma_{322}=m\cdot D' n',  \\
& \rho=\Gamma_{313}=m\cdot \delta n,\qquad \rho'=\Gamma_{323}=m\cdot \delta n',\qquad \beta=\Gamma_{213}=n'\cdot \delta n.\nn
\end{align}
Implicitly, one has $\beta'=-\beta$.

\noindent
Properties:
\begin{align}
& D n=-\kappa m + \epsilon n,\qquad D'n=- \tau m -\epsilon' n, \qquad \delta n=-\rho m + \beta n,\nn \\
&
Dn'=- \tau' m -\epsilon n',
\qquad D'n'=- \kappa' m + \epsilon' n', \qquad \delta n'=-\rho' m - \beta n', \\
& D m=- \tau' n -\kappa n' ,
\qquad D' m=- \kappa' n - \tau n' , \qquad \delta m=-\rho' n  - \rho n'.\nn
\end{align}
Geodesic deviation:
\be
\nabla_\mu n_\nu=-\kappa n'_\mu m_\nu + \epsilon n'_\mu n_\nu - \tau n_\mu m_\nu - \epsilon' n_\mu n_\nu + \rho m_\mu m_\nu - \beta m_\mu n_\nu.
\ee

\noindent{\bf NP equations}

\noindent
Curvature tensor:
\begin{align}
 D \epsilon' + D' \epsilon &=-2\epsilon \epsilon'+\beta (\tau'-\tau) + \tau \tau' - \kappa \kappa' + 2 (R_{12}-\frac14 R),\qquad [R_{1212}]\\
 D \beta - \delta \epsilon & =\beta \rho + \kappa \epsilon' + \kappa \rho' - \epsilon \tau'- \epsilon \beta - \rho \tau' - R_{13},\qquad [R_{1213}]\\
 D' \beta + \delta \epsilon' & = \tau \rho'+\epsilon'\tau + \beta \rho' - \beta \epsilon' -\epsilon \kappa'- \rho \kappa' + R_{23}, \qquad [R_{1223}]\\
D \rho - \delta \kappa & =\rho^2 + \epsilon \rho - 2 \kappa \beta - \kappa \tau - \kappa \tau' - R_{11},\qquad [R_{1313}]\\
D' \rho - \delta \tau & = \rho \rho' - \kappa\kappa' - \tau^2 - \rho \epsilon' + (R_{12}-\frac12 R), \qquad [R_{1323}]\\
D \rho' - \delta \tau' & = \rho \rho' - \kappa \kappa' - \tau'^2 - \epsilon \rho' + (R_{12}- \frac12 R), \qquad [R_{2313}]\\
D'\rho' - \delta \kappa' &= 2 \beta \kappa' + \rho'^2 - \tau \kappa' - \tau'\kappa'+\epsilon'\rho' - R_{22},\qquad [R_{2323}]\\
D\kappa' - D'\tau' & =\rho'\tau - 2 \kappa' \epsilon - \rho' \tau' + R_{23}, \qquad [R_{2312}]\\
D \tau - D' \kappa & =2\kappa \epsilon' + \rho \tau - \rho \tau' - R_{13}, \qquad [R_{1312}]
\end{align}

\noindent
Commutation relation:
\begin{align}
& D D' - D' D=-\epsilon D' + \epsilon' D + (\tau-\tau') \delta, \\
& D \delta - \delta D = - \kappa D' - (\beta + \tau') D + \rho \delta, \\
& D' \delta - \delta D' = - \kappa' D + (\beta - \tau) D' + \rho' \delta.
\end{align}

\subsection{ Bondi-Weyl gauge conditions and coordinates system}

In 4D NP formalism, a special coordinates system is constructed by introducing a family of null hypersurfaces labelled by the retarded time coordinate $u$ \cite{Newman:1961qr}, see also \cite{Newman:1962cia,Sachs:1962wk}. Then one null basis vector is chosen to be orthogonal to the null hypersurface. Since the hypersurface is null, the basis vector is tangent to a family of null geodesics on the hypersurface. It is convenient to choose $u$, the affine parameter $r$, and two other parameters that label the geodesics on each $u=$constant hypersurface as the coordinates.

Here we will follow a similar construction for the near horizon coordinates system. However, we will choose the advanced time coordinate $v$ which allows us to study both sides of the horizon \cite{Krishnan:2012bt}, since the retarded time coordinate $u$ does not cover the black hole region but the white hole region. The other two coordinates are the affine parameter $r$ and one remaining coordinate $\phi$ to label each geodesic. Following the 4D treatment \cite{Newman:1961qr,Newman:1962cia}, in this coordinates system, we can choose for the null basis
\be
n=-\frac{\p}{\p r}.
\ee
The minus sign indicates that $n$ is an ingoing null vector. Since $n$ is the normal vector of the $v=$constant null hypersurface, we should have that
\be\label{r}
n=W(v,r,\phi)dv.
\ee
This will yield the remaining two null basis vectors with the following forms
\be
n'=\frac{1}{W} \frac{\p}{\p v} + U \frac{\p}{\p r} - \frac{N}{W} \frac{\p}{\p \phi},\qquad m=\omega \frac{\p}{\p r} - \frac{1}{\Omega} \frac{\p}{\p \phi},
\ee
and
\be
n'=(UW+\omega \Omega N) dv - dr + \omega \Omega d \phi, \qquad m=\Omega(N dv + d \phi).
\ee
where $U,N,\omega,\Omega$ are arbitrary functions of the coordinates. We will further use the freedom from the rotation of the basis vectors to set $n'$ and $m$ parallelly propagated along $n$. In terms of the spin coefficients, those conditions yield
\be
\kappa=\epsilon=\tau'=0.
\ee
One can further use the freedom from the rotation of the basis vectors and a combined coordinates transformation to set $n$ being a gradient field as was chosen in \cite{Barnich:2016lyg}. However, we will not take this option in this work as it is in some sense more convenient to have exact solutions in this relaxed gauge. We refer to this gauge condition as Bondi-Weyl gauge following the terminology in \cite{Geiller:2021vpg}.

\subsection{ No black hole theorem}

\noindent{\bf In Bondi-Weyl gauge}

\noindent
The two components of the curvature tensor which are relevant to the no black hole theorem are
\begin{align}
D \rho - \delta \kappa & =\rho^2 + \epsilon \rho - 2 \kappa \beta - \kappa \tau - \kappa \tau' - R_{11},\\
D \rho' - \delta \tau' & = \rho \rho' - \kappa \kappa' - \tau'^2 - \epsilon \rho' + (R_{12}- \frac12 R).\label{main}
\end{align}
In the Bondi-Weyl gauge, they are reduced to
\begin{align}
D \rho  & =\rho^2 - R_{11},\\
D \rho' & = \rho \rho' + (R_{12}- \frac12 R).
\end{align}
We assume that there is an apparent horizon at $r=0$ and all the fields are given in a near horizon series expansion
\be
\phi= \sum_{a=0}^{\infty} \phi_a(v,\phi) r^a.
\ee
Moreover, this type of series expansion is valid on both sides of the horizon, namely $r>0$ outside the horizon while $r<0$ inside the horizon. An apparent horizon requires that $\rho_0>0$, $\rho'_0=0$, and $\rho'_1<0$. Einstein equation in NP formalism is in the form
\be
R_{\mu\nu}-\frac12 g_{\mu\nu} R + \Lambda g_{\mu\nu}=-T_{\mu\nu}.
\ee
Thus, it is clear that if $-R_{12}+ \frac12 R=T_{12}+\Lambda\geq0$, equation \eqref{main} yields that $\rho'_1\geq0$. This is in contradiction with the assumption that an apparent horizon is located at $r=0$. Hence, we can conclude that a black hole solution requires a negative cosmological constant if the stress tensor satisfies the dominant energy condition \cite{Ida:2000jh}.

\noindent{\bf In gauge conditions and coordinates system of \cite{Ida:2000jh}}

\noindent
In \cite{Ida:2000jh}, the no black hole theorem was obtained from a different gauge choice. In a 3D spacetime $(M,g)$, a spacelike hypersurface $\Sigma$ is selected. It is assumed that there is a trapped region on $\Sigma$ with an apparent horizon $H$ as the outer boundary of the trapped region on $H$. Then, the frame bases are chosen as follows. $m$ is a unit tangent vector of the apparent horizon $H$, $n$ and $n'$ are future directed outgoing and ingoing null vectors orthogonal to $H$. Note that we chose a different convention that $n$ is ingoing in previous discussion. Using boost transformation of $n$ and $n'$, it is possible to arrange that $n-n'$ lies in $\Sigma$. Moreover, it is assumed that $e^f(n-n')$ and $e^h m$ form holonomic bases on $\Sigma$ for some functions $f$ and $h$. Those requirements can be achieved in the coordinates $(t,r,\phi)$ with $\phi$ being periodic and $(r,\phi)$ spanning the spacelike hypersurface $\Sigma$. The conditions of holonomic bases yield that
\be
e^f(n-n')=\frac{\p}{\p r},\qquad e^h m=\frac{\p}{\p \phi}.
\ee
In terms of the spin coefficients, these conditions are indicating that
\be
\delta f=\kappa-\tau+\beta=\kappa'-\tau'-\beta.
\ee
Condition $n-n'$ lying in $\Sigma$ and the orthogonality conditions of the bases require that
\be
n-n'=-2 e^f dr, \qquad n+n' = 2 e^c dt,\qquad m = e^a dt - e^h d\phi,
\ee
and
\be
n+n' = e^{-c} \frac{\p}{\p t} + e^{-h} e^{a-c} \frac{\p}{\p \phi},
\ee
for some functions $a$ and $c$. Hence, the basis vectors are given by
\begin{align}
&n=\frac12 \left(e^{-c} \frac{\p}{\p t} + e^{-f} \frac{\p}{\p r} + e^{-h} e^{a-c} \frac{\p}{\p \phi} \right),\\
&n'=\frac12 \left(e^{-c} \frac{\p}{\p t} - e^{-f} \frac{\p}{\p r} + e^{-h} e^{a-c} \frac{\p}{\p \phi} \right),\\
&m=e^{-h} \frac{\p}{\p \phi}.
\end{align}
The no black hole theorem resides in the near horizon relation
\be
(D-D')\rho=\delta (\kappa-\tau) + (\kappa-\tau)^2 - R_{11}-(R_{12}-\frac12 R),
\ee
where we used the fact that $\rho=0$ at the horizon and the relation
\be
\kappa-\tau+\beta=\kappa'-\tau'-\beta.
\ee
A special choice of $f$ which sets $\delta(\delta f -\beta)=0$ leads to
\be
(D-D')\rho=(\kappa-\tau)^2 + T_{11} + T_{12} + \Lambda,
\ee
A positive cosmological constant and the dominant energy condition yield that the next-to-leading order of the expansion $\rho$ in the near horizon expansion is always positive, which is the statement of no black hole theorem \cite{Ida:2000jh}.

\subsection{ Solution space for standard Einstein gravity}

The near horizon solution space for 3D pure Einstein gravity with cosmological constant is first derived in \cite{Adami:2020ugu} in the form of near horizon expansion and in \cite{Geiller:2021vpg} with closed form, see also \cite{Adami:2021nnf,Adami:2022ktn,Geiller:2022vto} for several extensions. Here we present a self-contained derivation in the NP formalism. The solution of the spin coefficients from the radial equations in the Bondi-Weyl gauge is given by
\be
\begin{split}
& \rho = \frac{1}{r+A},\quad \rho_0=\frac{1}{A}, \quad \rho'=\frac{\rho'_0 A}{r+A} + \Lambda r - \frac{r^2}{2(r+A)}\Lambda, \\
&\beta = \frac{\beta_0 A}{r+A},\quad \tau=\frac{\tau_0 A}{r+A}, \quad \epsilon'=\beta_0\tau_0 A(1-\frac{A}{r+A}) +\epsilon'_0 - \Lambda r, \\
&\kappa'=\kappa'_0 - \frac12 A\tau_0\left[\Lambda r + \frac{A( \Lambda A - 2\rho'_0 )}{r+A} - \Lambda A + 2\rho'_0 \right],
\end{split}
\ee
where quantities with subscript $0$ are integration constants with respect to $r$. The commutation relations of the null bases lead to the following equations
\be
\begin{split}\label{commutator}
&D \frac{1}{W}=0,\quad D U=-\epsilon' + \omega \tau,\quad D \frac{N}{W}=\frac{\tau}{\Omega},\quad D\omega=\beta + \rho \omega,\quad D\frac{1}{\Omega}=\frac{\rho}{\Omega}, \\
& D'\omega-\delta U=\kappa'+(\beta-\tau)U + \rho' \omega,\quad -D'\frac{1}{\Omega} + \delta \frac{N}{W}=-(\beta-\tau)\frac{N}{W}-\frac{\rho'}{\Omega},\\
&-\delta \frac{1}{W}=(\beta-\tau)\frac{1}{W}.
\end{split}
\ee
The solutions of the first line in \eqref{commutator} are
\be
\begin{split}
&\Omega=\frac{\Omega_0}{A} (r+A),\qquad W=W_0,\\
&N=N_0 + \frac{W_0\tau_0 A}{\Omega_0} (\frac{A}{r+A}-1), \qquad \omega=\frac{\omega_0 A}{r+A}-\frac{\beta_0 A r}{r+A},\\
&U=-\frac12 \Lambda r^2 + (\beta_0\tau_0 A + \epsilon'_0)r + \frac{\omega_0 \tau_0 A^2 + \beta_0 \tau_0 A^3}{r+A} + U_0 - (\omega_0 \tau_0 A + \beta_0 \tau_0 A^2).
\end{split}
\ee
The solution to the last line of the commutation relation \eqref{commutator} is
\be
\tau_0 = \beta_0 - \frac{\p_\phi W_0}{\Omega_0 W_0}.
\ee
The second line of the commutation relation \eqref{commutator} and the remaining three relations from the curvature tensor determine only the time evolution of the integration constants $\beta_0$, $\rho_0$, $\rho'_0$, $\omega_0$, $\Omega_0$. The evolution relations are just tedious without further information about the solution space. We checked in Mathematica and will not present in full details here.


\section{3D charged rotating solution in NU gauge}
\label{3}

The equations of motion for Einstein-Maxwell theory in the convention of the NP formalism are
\be
R_{\mu\nu}-\frac12 g_{\mu\nu} R - \frac{1}{\ell^2} g_{\mu\nu}-\left(2F_{\mu\rho}{F_\nu}^\rho-\frac12 g_{\mu\nu} F^2\right)=0,\qquad \Lambda= - \frac{1}{\ell^2},
\ee
and
\be
\nabla_\nu F^{\mu\nu}=0.
\ee
The stress tensor of this case is
\be
T_{\mu\nu}=-\left(2F_{\mu\rho}{F_\nu}^\rho-\frac12 g_{\mu\nu} F^2\right).
\ee
The metric of the 3D charged rotating solution is \cite{Martinez:1999qi}
\be
ds^2=H^2F^2dT^2-\frac{dR^2}{F^2}-R^2(d\varphi-N^{\varphi}dT)^2,\\
\ee
and the gauge field is
\be
A=-\frac{Q\log r}{\sqrt{1-\frac{\omega^2}{\ell^2}}}\left(dT-\omega d\varphi \right),
\ee
where
\begin{align*}
& f^2=\frac{r^2}{\ell^2}-M-Q^2 \log r^2,\\
& R^2=\frac{r^2-f^2\omega^2}{1-\frac{\omega^2}{\ell^2}}=r^2+Y+X\log r^2,\quad Y=\frac{\omega^2 M}{1-\frac{\omega^2}{\ell^2}},\quad X=\frac{\omega^2 Q^2}{1-\frac{\omega^2}{\ell^2}},\\
& F^2=(\frac{dR}{dr})^2f^2, \quad H^2=\frac{r^2}{R^2}(\frac{dr}{dR})^2\\
& N^{\varphi}=\frac{\omega(\frac{r^2}{\ell^2}-f^2)}{(1-\frac{\omega^2}{\ell^2})R^2}=\frac{\omega(M+ Q^2\log r^2)}{(1-\frac{\omega^2}{\ell^2})R^2},
\end{align*}
where $\omega$, $M$ and $Q$ are constants which are related to angular velocity, mass, and charge parameters. In $(T,r,\varphi)$ coordinates, the metric is given by
\be
ds^2=\frac{r^2 f^2}{R^2}dT^2-\frac{dr^2}{f^2}-R^2(d\varphi-N^{\varphi}dT)^2.\\
\ee
Via coordinates transformation,
\bea
dv=dT+\frac{R dr}{r f^2},\;\;\;\;d\phi=d\varphi+\frac{N^{\varphi}R}{rf^2}dr,
\eea
the metric becomes
\be
ds^2=\frac{r^2 f^2}{R^2} dv^2 -2 \frac{r}{R} dvdr-R^2(d\phi - N^{\varphi}dv)^2,
\ee
and
\be\label{gauge}
A=-\frac{Q\log r}{\sqrt{1-\frac{\omega^2}{\ell^2}}}\left(dv-\omega d\phi \right)+\frac{Q\log r}{\sqrt{1-\frac{\omega^2}{\ell^2}}}(1-\omega N^\varphi)\frac{1}{HF^2}\frac{d R} {dr} d r.
\ee
The last term in the gauge field can be eliminated by a pure gauge transformation. Literally, this solution is not yet in the Bondi-Weyl gauge, because the metric element $g_{vr}$ still has radial dependence.\footnote{The radial dependence in the basis vector \eqref{r} is eliminated by commutation relation \eqref{commutator}. In the metric formalism, it is part of the gauge condition \cite{Adami:2020ugu,Geiller:2021vpg,Adami:2021nnf}.} This can be achieved by a pure radial coordinate transformation, namely defining $d\tilde{r}=\frac{r}{R} dr$. In the $(v,\tilde{r},\phi)$ coordinates, the 3D charged rotating solution is in the Bondi-Weyl gauge. Actually, the solution is even simpler, it is now written in the NU gauge.


\section{Horizon of 3D charged rotating solution}
\label{4}

The apparent horizon is defined to be the outermost marginally trapped surface. Let us choose two null directions $n'$ being outgoing and $n$ being incoming following the choice of the Bondi-Weyl gauge,
\be
n'=\frac12 \frac{r^2 f^2}{R^2} dv - d\tilde{r},\qquad n= dv.
\ee
The expansions of the two null directions can be computed directly. They are given by
\begin{align}
&\rho'=-\frac{r^2 f^2 \frac{d R}{d \tilde {r}}}{2 R^3}=-\frac{r^2 f^2 \frac{d R}{d r}\frac{dr}{d \tilde{r}}}{2 R^3}=-\frac{r^2 f^2 }{2 R^2}\frac{d R}{d r} \frac{1}{r},\\
&\rho=\frac{\frac{d R}{d \tilde {r}}}{ R}=\frac{\frac{d R}{d r}\frac{dr}{d\tilde{r}}}{ R}=\frac{d R}{d r}\frac{1}{r}.
\end{align}
Since $\rho'$ and $\rho$ are scalar fields, they are given by the same value in different coordinates. So we will use coordinate $r$ to characterize the location of the horizon which is much simpler. The property of the horizon depends on the cosmological constant. We will discuss that separately for negative, zero, and positive cosmological constant.

\subsection{Negative cosmological constant}

For the negative cosmological constant case, it is proven that the 3D charged rotating solution is a black hole solution for some particular regions of the mass, charge, and angular velocity parameters \cite{Martinez:1999qi}. For completeness, we will present a self-contained discussion on this issue which will be useful for the discussions of other choices of the cosmological constant. The expansions of the two null directions now are
\begin{align}
&\rho'=-\frac{r^2 f^2 }{2 R^2}\frac{d R}{d r} \frac{1}{r}=\frac{ 1 }{2 R^3}(r^2+\frac{\omega^2Q^2}{1-\frac{\omega^2}{\ell^2}})(-\frac{r^2}{\ell^2}+M+Q^2 \log r^2),\\
&\rho=\frac{1}{r^2 R} (r^2+\frac{\omega^2Q^2}{1-\frac{\omega^2}{\ell^2}}).
\end{align}
Those expressions involve
\be
R=\sqrt{r^2+ \frac{\omega^2}{1-\frac{\omega^2}{\ell^2}} (M+Q^2 \log r^2)},
\ee
a square root function, in the denominator, so it must be positive. Alternatively, this condition defines the region where the coordinate $r$ is valid for describing the spacetime. We see that $R^2$ is a monotonically increasing function of $r$.
Suppose that $r=r_0$ is the only root of $R^2=0$, we should restrict ourselves in the region $r>r_0$. The expression of the gauge field in \eqref{gauge} requires that $1-\frac{\omega^2}{\ell^2}>0$. The reason for such a requirement is that the rotating parameter $\omega$ is introduced from a boost transformation with such assumption \cite{Martinez:1999qi}. We will further comment on the boost transformation in the next section. The function $-f^2=-\frac{r^2}{\ell^2}+M+Q^2 \log r^2$ has the maximum value at $r_m=Q^2\ell^2$. To guarantee that $-f^2=0$ has a root, the mass and charge parameter must satisfy the condition
\be
- Q^2+M+Q^2\log (Q^2\ell^2)>0.
\ee
Then $-f^2=0$ will have two roots. Supposing that $r=r_i$ and $r=r_h$ are the smaller and larger roots respectively, it is clear that $r_0<r_i<r_h$. When $r>r_h$, we have $\rho'<0$ and $\rho>0$, it is the normal region. While, both expansions $\rho$ and $\rho'$ are larger than zero between the two horizons, namely $r_i<r<r_h$. This is the black hole region and $r=r_h$ defines the apparent horizon of the black hole. Correspondingly, the stress tensor should satisfy that $T_{12}-\frac{1}{\ell^2}<0$. The $T_{12}$ component of the stress tensor is
\be
T_{12}=\frac{Q^2(\omega N^\varphi - 1)^2 \frac{dr}{d \tilde {r}}}{(1-\frac{\omega^2}{\ell^2})r^2}=\frac{Q^2}{(1-\frac{\omega^2}{\ell^2})R^2}.
\ee
On the horizon it reduces to
\be
T_{12}=\frac{Q^2}{r_h^2}.
\ee
The fact $r_h^2>r_m^2=Q^2\ell^2$ ensures that $T_{12}-\frac{1}{\ell^2}<0$.

\subsection{Zero cosmological constant}

In the $\Lambda=0$ case, substituting $\ell=\infty$ into the 3D charged rotating solution, the expansions of the null vectors $n'$, $n$ are given by
\begin{align}
&\rho'=-\frac{r^2 f^2 }{2 R^2}\frac{d R}{d r} \frac{1}{r}=\frac{ 1 }{2 R^3}(r^2+ \omega^2Q^2 )(M+Q^2\log{r^2}),\\
&\rho=\frac{1}{r^2 R} (r^2+\omega^2Q^2).
\end{align}
Suppose that $r=r_0$ is the root of $R^2=r^2 + \omega^2(M+Q^2\log r^2)=0$, we should restrict ourselves in the region $r>r_0$. There is only one root from $M+Q^2\log{r^2}=0$, which is located at $r_h=e^{-\frac{M}{2Q^2}}$ and it is obvious that $r_0<r_h$. This is a cosmological horizon as both expansions $\rho$ and $\rho'$ are larger than zero when $r>r_h$. Hence this is the trapped region of the spacetime. While the spacetime is regular when $r_0<r<r_h$. This is in consistency with the no black hole theorem presented in previous section. Correspondingly, the $T_{12}$ component of the stress tensor is larger than zero at $r_h$,
\be
T_{12}=\frac{Q^2}{R^2}\bigg|_{r=r_h}.
\ee
The presence of Maxwell fields does exclude a black hole formation in the case of $\Lambda=0$ \cite{Ida:2000jh}.

\subsection{Positive cosmological constant}

In the case of a positive cosmological constant, all $\ell^2$ terms should be replaced by $-\ell^2$, and the expansions of null vectors $n'$ and $n$ can be directly calculated as
\begin{align}
&\rho'=-\frac{r^2 f^2 }{2 R^2}\frac{d R}{d r} \frac{1}{r}=\frac{ 1 }{2 R^3}(r^2+\frac{\omega^2Q^2}{1+\frac{\omega^2}{\ell^2}})(\frac{r^2}{\ell^2}+ M+Q^2\log{r^2}),\\
&\rho=\frac{1}{r^2 R} (r^2+\frac{\omega^2Q^2}{1+\frac{\omega^2}{\ell^2}}).
\end{align}
Since both $R^2=r^2+ \frac{\omega^2}{1+\frac{\omega^2}{\ell^2}} (M+Q^2 \log r^2)$ and $-f^2=\frac{r^2}{\ell^2}+ M+Q^2\log{r^2}$ are monotonically increasing functions of $r$, they both can only have one root. Suppose that $r=r_0$ and $r=r_h$ are the roots of the two functions respectively, it is obvious that $r_0<r_h$. Both expansions are positive when $r>r_h$, while $\rho>0$ and $\rho'<0$ in the region $r_0<r<r_h$. So $r_h$ is a cosmoligical horizon which is similar to the $\Lambda=0$ case. The $T_{12}$ component of the stress tensor is larger than zero at $r_h$,
\be
T_{12}=\frac{Q^2}{(1+\frac{\omega^2}{\ell^2})R^2}\bigg|_{r=r_h}.
\ee


\section{Comment on flat limit of 3D charged rotating solution}
\label{5}

The 3D black hole of standard Einstein gravity with a negative cosmological constant arises from identification of points of AdS$_3$ spacetime. The static and rotating black hole are obtained from identification along different directions \cite{Banados:1992gq}. The static BTZ black hole with the metric
\be
ds^2=-f^2 dt^2 + \frac{dr^2}{f^2} + r^2 d\theta^2,\quad f^2=\frac{r^2}{\ell^2}-m,
\ee
arises from the identification $\theta\rightarrow \theta + 2 k \pi $ \cite{Banados:1992gq}, where $k$ is an integer. The rotating BTZ black hole
\be\label{btz}
ds^2=-N^2 dT^2 + \frac{dR^2}{N^2} + R^2( d\varphi - N^\varphi dT)^2,
\ee
where
\be
N^2=\frac{R^2}{\ell^2} - M + \frac{J^2}{4R^2},\quad N^\varphi=\frac{J}{2R^2}.
\ee
arises from the identification $\varphi\rightarrow \varphi + 2 k \pi $ \cite{Banados:1992gq}. Before the identification, the two solutions are both describing AdS$_3$ spacetime and they are connected by a boost transformation \cite{Martinez:1999qi}, see also \cite{Deser:1983tn,Deser:1985pk,Brown:1988am,Clement:1995zt},
\be\label{boost}
t=\frac{T-\omega \varphi}{\sqrt{1-\frac{\omega^2}{\ell^2}}},\quad \theta=\frac{\varphi-\frac{\omega}{\ell^2} T}{\sqrt{1-\frac{\omega^2}{\ell^2}}}.
\ee
In the new coordinates $(T,R,\varphi)$, it is the metric of the rotating BTZ \eqref{btz} with the following identifications,
\be
R^2=\frac{r^2-f^2\omega^2}{1-\frac{\omega^2}{\ell^2}},\qquad M=m\frac{1+\frac{\omega^2}{\ell^2}}{1-\frac{\omega^2}{\ell^2}},\qquad J=\frac{2\omega m}{1-\frac{\omega^2}{\ell^2}}.
\ee
The boost transformation can be implemented as a generic trick to create rotating solutions from static ones in 3D. In practice, one should first unwrap the identification of the static black hole solution. The second step is to apply the boost transformation to obtain the rotating metric before identification. The last step is to choose a new Killing direction to identify points to recover the rotating black hole. This is precisely the way that the authors of Ref. \cite{Martinez:1999qi} found the charged rotating black hole.

When taking the flat limit $\ell\rightarrow\infty$, it is obvious from the boost transformation \eqref{boost} that $\theta=\varphi$. Hence, the two solutions are obtained from identification along the same direction. They are just ordinary coordinates transformation of each other, see, for instance, in \cite{Barnich:2012aw} for more details of the non-charged case. For the charged case, it is the same boost transformation to build the rotating black hole in \cite{Martinez:1999qi}. Thus, the rotating parameter of the flat limit of the 3D charged rotating solution is again introduced from pure coordinates transformation which does not describe an intrinsic rotation of the source.


\section{Conclusion}

In this paper, we present an alternative derivation of the 3D no black hole theorem in the NP formalism. We show that, in the well established NU gauge choice, the no black hole theorem is transparent from the NP equations. The existence of an apparent horizon requires a negative cosmological constant. The near horizon solution space of pure Einstein gravity is derived in a self-contained way in the NP formalism. The solution space is in a closed form which is consistent with previous results \cite{Geiller:2021vpg,Adami:2022ktn}. As a consistency check, we study in detailed the horizon property of the 3D charged rotating solution \cite{Clement:1993kc,Martinez:1999qi}. A black hole solution only exists in the case of a negative cosmological constant. For the cases of vanishing or positive cosmological constant, there will be a cosmological horizon. We examine the flat limit of the 3D charged rotating solution and show that the spinning parameter is from a pure coordinates transformation which is the same as the non-charged case \cite{Barnich:2012aw}.

\section*{Acknowledgments}
The authors thank Xiaoning Wu and Run-Qiu Yang for valuable discussions. The authors thank Professor G. Clement for informing us that the 3D charged rotating solution was first derived in \cite{Clement:1993kc}. This work is supported in part by the National Natural Science Foundation of China (NSFC) under Grants No. 11905156 and No. 11935009.

\providecommand{\href}[2]{#2}\begingroup\raggedright\endgroup


\providecommand{\href}[2]{#2}\begingroup\raggedright\begin{thebibliography}{10}

\bibitem{Banados:1992wn}
M.~Banados, C.~Teitelboim, and J.~Zanelli, ``{The Black hole in
  three-dimensional space-time},''
  \href{http://dx.doi.org/10.1103/PhysRevLett.69.1849}{{\em Phys. Rev. Lett.}
  {\bfseries 69} (1992) 1849--1851},
  \href{http://arxiv.org/abs/hep-th/9204099}{{\ttfamily arXiv:hep-th/9204099}}.

\bibitem{Brown:1986nw}
J.~D. Brown and M.~Henneaux, ``{Central Charges in the Canonical Realization of
  Asymptotic Symmetries: An Example from Three-Dimensional Gravity},''
  \href{http://dx.doi.org/10.1007/BF01211590}{{\em Commun. Math. Phys.}
  {\bfseries 104} (1986) 207--226}.

\bibitem{Strominger:1997eq}
A.~Strominger, ``{Black hole entropy from near horizon microstates},''
  \href{http://dx.doi.org/10.1088/1126-6708/1998/02/009}{{\em JHEP} {\bfseries
  02} (1998) 009}, \href{http://arxiv.org/abs/hep-th/9712251}{{\ttfamily
  arXiv:hep-th/9712251}}.

\bibitem{Aharony:1999ti}
O.~Aharony, S.~S. Gubser, J.~M. Maldacena, H.~Ooguri, and Y.~Oz, ``{Large N
  field theories, string theory and gravity},''
  \href{http://dx.doi.org/10.1016/S0370-1573(99)00083-6}{{\em Phys. Rept.}
  {\bfseries 323} (2000) 183--386},
  \href{http://arxiv.org/abs/hep-th/9905111}{{\ttfamily arXiv:hep-th/9905111}}.

\bibitem{Banados:1992gq}
M.~Banados, M.~Henneaux, C.~Teitelboim, and J.~Zanelli, ``{Geometry of the
  (2+1) black hole},'' \href{http://dx.doi.org/10.1103/PhysRevD.48.1506}{{\em
  Phys. Rev. D} {\bfseries 48} (1993) 1506--1525},
  \href{http://arxiv.org/abs/gr-qc/9302012}{{\ttfamily arXiv:gr-qc/9302012}}.
  [Erratum: Phys.Rev.D 88, 069902 (2013)].

\bibitem{Ida:2000jh}
D.~Ida, ``{No black hole theorem in three-dimensional gravity},''
  \href{http://dx.doi.org/10.1103/PhysRevLett.85.3758}{{\em Phys. Rev. Lett.}
  {\bfseries 85} (2000) 3758--3760},
  \href{http://arxiv.org/abs/gr-qc/0005129}{{\ttfamily arXiv:gr-qc/0005129}}.

\bibitem{Newman:1961qr}
E.~Newman and R.~Penrose, ``{An Approach to gravitational radiation by a method
  of spin coefficients},'' \href{http://dx.doi.org/10.1063/1.1724257}{{\em J.
  Math. Phys.} {\bfseries 3} (1962) 566--578}.

\bibitem{Newman:1962cia}
E.~T. Newman and T.~W.~J. Unti, ``{Behavior of Asymptotically Flat Empty
  Spaces},'' \href{http://dx.doi.org/10.1063/1.1724303}{{\em J. Math. Phys.}
  {\bfseries 3} no.~5, (1962) 891}.

\bibitem{Clement:1993kc}
G.~Clement, ``{Classical solutions in three-dimensional Einstein-Maxwell
  cosmological gravity},''
  \href{http://dx.doi.org/10.1088/0264-9381/10/5/002}{{\em Class. Quant. Grav.}
  {\bfseries 10} (1993) L49--L54}.

\bibitem{Martinez:1999qi}
C.~Martinez, C.~Teitelboim, and J.~Zanelli, ``{Charged rotating black hole in
  three space-time dimensions},''
  \href{http://dx.doi.org/10.1103/PhysRevD.61.104013}{{\em Phys. Rev. D}
  {\bfseries 61} (2000) 104013},
  \href{http://arxiv.org/abs/hep-th/9912259}{{\ttfamily arXiv:hep-th/9912259}}.

\bibitem{Barnich:2016lyg}
G.~Barnich and C.~Troessaert, ``{Finite BMS transformations},''
  \href{http://dx.doi.org/10.1007/JHEP03(2016)167}{{\em JHEP} {\bfseries 03}
  (2016) 167}, \href{http://arxiv.org/abs/1601.04090}{{\ttfamily
  arXiv:1601.04090 [gr-qc]}}.

\bibitem{Sachs:1962wk}
R.~K. Sachs, ``{Gravitational waves in general relativity. 8. Waves in
  asymptotically flat space-times},''
  \href{http://dx.doi.org/10.1098/rspa.1962.0206}{{\em Proc. Roy. Soc. Lond. A}
  {\bfseries 270} (1962) 103--126}.

\bibitem{Krishnan:2012bt}
B.~Krishnan, ``{The spacetime in the neighborhood of a general isolated black
  hole},'' \href{http://dx.doi.org/10.1088/0264-9381/29/20/205006}{{\em Class.
  Quant. Grav.} {\bfseries 29} (2012) 205006},
  \href{http://arxiv.org/abs/1204.4345}{{\ttfamily arXiv:1204.4345 [gr-qc]}}.

\bibitem{Geiller:2021vpg}
M.~Geiller, C.~Goeller, and C.~Zwikel, ``{3d gravity in Bondi-Weyl gauge:
  charges, corners, and integrability},''
  \href{http://dx.doi.org/10.1007/JHEP09(2021)029}{{\em JHEP} {\bfseries 09}
  (2021) 029}, \href{http://arxiv.org/abs/2107.01073}{{\ttfamily
  arXiv:2107.01073 [hep-th]}}.

\bibitem{Adami:2020ugu}
H.~Adami, M.~M. Sheikh-Jabbari, V.~Taghiloo, H.~Yavartanoo, and C.~Zwikel,
  ``{Symmetries at null boundaries: two and three dimensional gravity cases},''
  \href{http://dx.doi.org/10.1007/JHEP10(2020)107}{{\em JHEP} {\bfseries 10}
  (2020) 107}, \href{http://arxiv.org/abs/2007.12759}{{\ttfamily
  arXiv:2007.12759 [hep-th]}}.

\bibitem{Adami:2021nnf}
H.~Adami, D.~Grumiller, M.~M. Sheikh-Jabbari, V.~Taghiloo, H.~Yavartanoo, and
  C.~Zwikel, ``{Null boundary phase space: slicings, news \& memory},''
  \href{http://dx.doi.org/10.1007/JHEP11(2021)155}{{\em JHEP} {\bfseries 11}
  (2021) 155}, \href{http://arxiv.org/abs/2110.04218}{{\ttfamily
  arXiv:2110.04218 [hep-th]}}.

\bibitem{Adami:2022ktn}
H.~Adami, P.~Mao, M.~M. Sheikh-Jabbari, V.~Taghiloo, and H.~Yavartanoo,
  ``{Symmetries at causal boundaries in 2D and 3D gravity},''
  \href{http://dx.doi.org/10.1007/JHEP05(2022)189}{{\em JHEP} {\bfseries 05}
  (2022) 189}, \href{http://arxiv.org/abs/2202.12129}{{\ttfamily
  arXiv:2202.12129 [hep-th]}}.

\bibitem{Geiller:2022vto}
M.~Geiller and C.~Zwikel, ``{The partial Bondi gauge: Further enlarging the
  asymptotic structure of gravity},''
  \href{http://dx.doi.org/10.21468/SciPostPhys.13.5.108}{{\em SciPost Phys.}
  {\bfseries 13} (2022) 108}, \href{http://arxiv.org/abs/2205.11401}{{\ttfamily
  arXiv:2205.11401 [hep-th]}}.

\bibitem{Deser:1983tn}
S.~Deser, R.~Jackiw, and G.~'t~Hooft, ``{Three-Dimensional Einstein Gravity:
  Dynamics of Flat Space},''
  \href{http://dx.doi.org/10.1016/0003-4916(84)90085-X}{{\em Annals Phys.}
  {\bfseries 152} (1984) 220}.

\bibitem{Deser:1985pk}
S.~Deser and P.~O. Mazur, ``{Static Solutions in $D=3$ Einstein-maxwell
  Theory},'' \href{http://dx.doi.org/10.1088/0264-9381/2/3/003}{{\em Class.
  Quant. Grav.} {\bfseries 2} (1985) L51}.

\bibitem{Brown:1988am}
J.~D. Brown, {\em {LOWER DIMENSIONAL GRAVITY}}.
\newblock 1988.

\bibitem{Clement:1995zt}
G.~Clement, ``{Spinning charged BTZ black holes and selfdual particle - like
  solutions},'' \href{http://dx.doi.org/10.1016/0370-2693(95)01464-0}{{\em
  Phys. Lett. B} {\bfseries 367} (1996) 70--74},
  \href{http://arxiv.org/abs/gr-qc/9510025}{{\ttfamily arXiv:gr-qc/9510025}}.

\bibitem{Barnich:2012aw}
G.~Barnich, A.~Gomberoff, and H.~A. Gonzalez, ``{The Flat limit of three
  dimensional asymptotically anti-de Sitter spacetimes},''
  \href{http://dx.doi.org/10.1103/PhysRevD.86.024020}{{\em Phys. Rev. D}
  {\bfseries 86} (2012) 024020},
  \href{http://arxiv.org/abs/1204.3288}{{\ttfamily arXiv:1204.3288 [gr-qc]}}.

\end{thebibliography}\endgroup
\end{document}